\documentclass[preprint,superscriptaddress,nofootinbib,onecolumn,aps,prd]{revtex4-1}
\pdfoutput=1

\usepackage[utf8]{inputenc}
\usepackage{amsmath,amssymb}
\usepackage{array}

\begin{document}

\title{Suppression of $H\to VV$ decay channels in the Georgi-Machacek model}

\author{S.I. Godunov}
\email{sgodunov@itep.ru}
\affiliation{Institute for Theoretical and Experimental Physics, Moscow, 117218, Russia} 
\affiliation{Novosibirsk State University, Novosibirsk, 630090, Russia} 

\author{M.I. Vysotsky}
\email{vysotsky@itep.ru}
\affiliation{Institute for Theoretical and Experimental Physics, Moscow, 117218, Russia} 
\affiliation{Moscow Institute of Physics and Technology, 141700, Dolgoprudny, Moscow 
Region, Russia}
\affiliation{Moscow Engineering Physics Institute, 115409, Moscow, Russia}

\author{E.V. Zhemchugov}
\email{zhemchugov@itep.ru}
\affiliation{Institute for Theoretical and Experimental Physics,
  Moscow, 117218, Russia}
\affiliation{Moscow Engineering Physics Institute, 115409, Moscow, Russia}

\begin{abstract}
  The $H\to ZZ$ decay mode is usually considered as one of the most
  promising ways to discover new heavy neutral scalar $H$. We show
  that in the Georgi--Machacek model it is possible to get large
  enhancement of double SM-like Higgs boson production due to $H$
  decays while $ZZ$ and $WW$ decay channels could be highly
  suppressed.
\end{abstract}

\maketitle

\section{Introduction}
\label{sec:introduction}

Models with extended Higgs sector provide a very rich
phenomenology. In many models there is a heavy neutral scalar $H$
which has the admixture of the neutral component of the Standard Model
(SM) doublet. As one of the consequences this new heavy scalar can
provide us with the significant enhancement of double SM-like Higgs
boson ($h$) production.

In our previous paper \cite{GVZ2014} we considered extensions of the
SM by isotriplets with hypercharges $Y=0$ and $Y=2$. In see-saw type
II (one extra triplet with $Y=2$) double Higgs boson production can be
increased by the value which is comparable with what we have in the
SM. In the Georgi--Machacek (GM) model \cite{Georgi} we obtained that
$H$ production cross section is significantly enhanced so $2h$
production can be much larger than in the SM.

If $H$ is produced with large cross section then in principle it can
be discovered in $ZZ$ final state. Direct searches in this mode at the
LHC \cite{ZZsearch} can set limits on model parameters which in some
ranges of $M_{H}$ can be stronger than that from $h$
couplings. However it is not the case in the GM model since we show
that ${\rm Br}\left(H\to ZZ\right)<1\%$ could occur.

\section{The model}
\label{sec:model}

In the GM model (see \cite{Logan2014} for a detailed review of this
model\footnote{We follow the notations used in \cite{Logan2014} with
  minor change: $\chi\leftrightarrow\Delta$, $v_{\chi}\leftrightarrow
  v_{\Delta}$.}) in addition to SM Higgs doublet
\begin{equation}
  \label{eq:doublet}
  \Phi^{(0)}=
  \begin{bmatrix}
    \Phi^{+}\\
    \Phi^{0}
  \end{bmatrix},
\end{equation}
two isotriplets with hypercharges $Y=0$ and $Y=2$ are introduced:
\begin{equation}
  \label{eq:triptets}
  \xi=
  \begin{bmatrix}
    \xi^{+}\\
    \xi^{0}\\
    \xi^{-}
  \end{bmatrix},\hspace{1cm}
  \Delta=
  \begin{bmatrix}
    \Delta^{++}\\
    \Delta^{+}\\
    \Delta^{0}
  \end{bmatrix}.
\end{equation}

We took the potential in the following form (for a detailed
description see \cite{Logan2014}):
\begin{equation}
  \label{eq:potential}
  V=\frac{\mu_{2}^{2}}{2}{\rm Tr}\left(\Phi^{\dagger}\Phi\right)
    +\lambda_{1}\left[{\rm Tr}\left(\Phi^{\dagger}\Phi\right)\right]^{2}
    +\frac{\mu_{3}^{2}}{2}{\rm Tr}\left(X^{\dagger}X\right)
    -M_{1}{\rm Tr}\left(\Phi^{\dagger}\tau^{a}\Phi\tau^{b}\right)\left(UXU^{\dagger}\right)_{ab},
\end{equation}
where $\Phi^{(0)}$, $\xi$ and $\Delta$ are combined into matrices
$\Phi$ and $X$:
\begin{eqnarray}
  \label{eq:matrices}
  \Phi=
  \begin{bmatrix}
    \Phi^{0*}&\Phi^{+}\\
    -\Phi^{+*}&\Phi^{0}
  \end{bmatrix},
  &\hspace{1cm}\langle\Phi\rangle=\frac{1}{\sqrt{2}}
  \begin{bmatrix}
    v_{\phi}&0\\
    0&v_{\phi}
  \end{bmatrix},\\
  X=
  \begin{bmatrix}
    \Delta^{0*}&\xi^{+}&\Delta^{++}\\
    -\Delta^{+*}&\xi^{0}&\Delta^{+}\\
    \Delta^{++*}&-\xi^{+*}&\Delta^{0}
  \end{bmatrix},
  &\hspace{1cm}\langle X\rangle=
  \begin{bmatrix}
    v_{\Delta}&0&0\\
    0&v_{\Delta}&0\\
    0&0&v_{\Delta}
  \end{bmatrix},
\end{eqnarray}
$U$ is a rotational matrix:
\begin{equation}
  \label{eq:U}
  U=
  \begin{bmatrix}
    -\frac{1}{\sqrt{2}}&0&\frac{1}{\sqrt{2}}\\
    -\frac{i}{\sqrt{2}}&0&-\frac{i}{\sqrt{2}}\\
    0&1&0
  \end{bmatrix},
\end{equation}
$\tau^{a}=\sigma^{a}/2$ where $\sigma^{a}$ are Pauli matrices.  Let us
note that we consider simplified potential which corresponds to
$\lambda_{2},\lambda_{3},\lambda_{4},\lambda_{5},M_{2}=0$ choice in
the potential considered in paper \cite{Logan2014}.

The value of $v_{\Delta}$ is generated by the term proportional to
$M_{1}$ and the following relations are useful:
\begin{equation}
  \label{eq:M1}
  v_{\Delta}=\frac{M_{1}v_{\Phi}^{2}}{4\mu_{3}^{2}},\hspace{1cm}
  v_{\phi}^{2}+8v_{\Delta}^{2}=\left(246~\mbox{GeV}\right)^{2}.
\end{equation}
In paper \cite{GVZ2014} $v_{\Delta}$ is defined to be $\sqrt{2}$ times
larger.

Since the potential is written in the way that vevs of two triplets
are the same then the custodial symmetry is preserved and the relation
between $W$ and $Z$ boson masses is the same as in the SM so the
experimental data on $W$ mass does not lead to new limits on model
parameters. It means that main restrictions originate from the
measurement of $h$ couplings to SM particles. Since the accuracy of
these measurements is not very good at the moment, these restrictions
are not very tough and $v_{\Delta}$ up to approximately
$30~\mbox{GeV}$ is allowed.

Only one scalar, a combination of $\xi^{0}$ and $\Delta^{0}$, mixes
with neutral component $\Phi^{0}$ of SM doublet forming mass
eigenstates, $h$ and $H$, which correspond to the scalar discovered at
LHC (so $M_{h}=125~\mbox{GeV}$) and new heavy scalar. Since this new
scalar $H$ has doublet admixture, it couples to quarks and therefore
it can be produced in gluon-gluon fusion so its production cross
section at LHC can be much larger than that for the other scalars of
the GM model which can be produced only in electroweak processes. It
was stressed in \cite{GVZ2014} that in some region of parameters space
$H$ decays can provide great enhancement of double $h$ production at
LHC.

\section{Heavy scalar decays}
\label{sec:decays}

According to papers \cite{Logan2014,Yagyu2} the couplings of $H$ boson
to $hh$, $WW$, and $ZZ$ are the following:
\begin{eqnarray}
  \label{eq:couplings_hh}
  g_{Hhh}=24\lambda_{1}c_{\alpha}^{2}s_{\alpha}v_{\phi}-\frac{\sqrt{3}}{2}M_{1}c_{\alpha}\left(3c_{\alpha}^{2}-2\right),\\
  \label{eq:couplings_VV}
  g_{HWW}=c_{W}^{2}g_{HZZ}=\frac{g^{2}}{6}\left(8\sqrt{3}c_{\alpha}v_{\Delta}+3s_{\alpha}v_{\phi}\right),
\end{eqnarray}
where $c_{W}=\cos\Theta_{W}$, $\Theta_{W}$ is the weak mixing angle,
$c_{\alpha}=\cos\alpha$, $s_{\alpha}=\sin\alpha$, $\alpha$ is the
mixing angle between $h$ and $H$.

According to the potential (\ref{eq:potential}) $\alpha$ is defined as
following:
\begin{equation}
  \label{eq:mixing_angle}
  \sin 2\alpha=\frac{-\sqrt{3}v_{\phi}M_{1}}{M_{H}^{2}-M_{h}^{2}},
\end{equation}
where $M_{H}$ is the mass of $H$.

Using coupling constants (\ref{eq:couplings_hh}) and
(\ref{eq:couplings_VV}), for the partial widths of $H$ decays we get
\begin{align}
  \label{eq:branchings_hh}
  \Gamma_{H\to hh}\approx&\frac{v_{\Delta}^{2}}{v_{\phi}^{4}}\frac{3M_{H}^{3}}{8\pi}
  \left[\frac{1+2\left(\frac{M_{h}}{M_{H}}\right)^{2}}{1-\left(\frac{M_{h}}{M_{H}}\right)^{2}}\right]^{2}
  \sqrt{1-4\frac{M_{h}^{2}}{M_{H}^{2}}},&\\
  \label{eq:branchings_ZZ}
  \Gamma_{H\to ZZ}\approx&\frac{v_{\Delta}^{2}}{v_{\phi}^{4}}\frac{M_{H}^{3}}{24\pi}
  \left[\frac{1-4\left(\frac{M_{h}}{M_{H}}\right)^{2}}{1-\left(\frac{M_{h}}{M_{H}}\right)^{2}}\right]^{2}
  \left(1-4\frac{M_{Z}^{2}}{M_{H}^{2}}+12\frac{M_{Z}^{4}}{M_{H}^{4}}\right)
  \sqrt{1-4\frac{M_{Z}^{2}}{M_{H}^{2}}},&\\
  \label{eq:branchings_WW}
  \Gamma_{H\to WW}\approx&\frac{v_{\Delta}^{2}}{v_{\phi}^{4}}\frac{M_{H}^{3}}{12\pi}
  \left[\frac{1-4\left(\frac{M_{h}}{M_{H}}\right)^{2}}{1-\left(\frac{M_{h}}{M_{H}}\right)^{2}}\right]^{2}
  \left(1-4\frac{M_{W}^{2}}{M_{H}^{2}}+12\frac{M_{W}^{4}}{M_{H}^{4}}\right)
  \sqrt{1-4\frac{M_{W}^{2}}{M_{H}^{2}}}.&
\end{align}
Deriving these formulae we used the approximation $v_{\phi}\gg
v_{\Delta}$, i.e. $\sin2\alpha\approx2\sin\alpha$, $\mu_{3}\approx
M_{H}$, and $8\lambda_{1}v_{\phi}^{2}\approx M_{h}^{2}$. For
$v_{\Delta}=20~\mbox{GeV}$ we get $\sin\alpha=0.35$.

In paper \cite{GVZ2014} it was found that $H\to hh$ decays can provide
large enhancement of double Higgs boson production. For
$M_{H}=300~\mbox{GeV}$ and $v_{\Delta}\approx 20~\mbox{GeV}$ we get
$\sigma(gg\to H)=\sin^{2}\alpha\times\sigma\left(gg\to H^{\rm
    SM}\right)\approx 1.4~\mbox{pb}$ at
$\sqrt{s}=14~\mbox{TeV}$. Using (\ref{eq:branchings_hh}),
(\ref{eq:branchings_ZZ}), and (\ref{eq:branchings_WW}) for
$M_{H}=300~\mbox{GeV}$ we get ${\rm Br}\left(H\to hh\right)\approx
98\%$ while ${\rm Br}\left(H\to ZZ\right)\approx 0.6\%$. It means that
in spite of large $H$ production cross section the enhancement in $ZZ$
final state is negligible so the search for $H$ in this mode at LHC
will not lead to new limits on model parameters.

\section{Conclusions}
\label{sec:conclusions}

It was shown that though in the GM model new heavy neutral scalar $H$
can be produced with large cross section at the LHC, $ZZ$ and $WW$
decay modes can be very suppressed (if $H\to hh$ decays are
kinematically allowed and $M_{H}$ is not significantly larger than
$300~\mbox{GeV}$) so direct searches for $H$ in these decay modes will
not lead to its discovery. This is a peculiar feature of the GM model.

The authors are partially supported under the grants RFBR
No. 14-02-00995 and NSh-3830.2014.2. S G. and E. Zh. are also
supported by MK-4234.2015.2. In addition, S. G is supported by Dynasty
Foundation and by the Russian Federation Government under grant
No. 11.G34.31.0047.

\end{document}